\documentclass{ifacconf}
\usepackage{amssymb, amsmath}
\usepackage{mathtools}
\usepackage{autobreak}
\usepackage{float}
\usepackage{graphicx} 
\usepackage{natbib} 
\usepackage{amsfonts}
\usepackage{mathrsfs}
\usepackage{color}
\usepackage{caption}
\usepackage{subcaption}

\newcommand{\KeyGen}{\mathsf{Gen}}
\newcommand{\Ecd}{\mathsf{Ecd}}
\newcommand{\Dcd}{\mathsf{Dcd}}
\newcommand{\Enc}{\mathsf{Enc}}
\newcommand{\Dec}{\mathsf{Dec}}
\newcommand{\pk}{\mathsf{pk}}
\newcommand{\sk}{\mathsf{sk}}
\newcommand{\bMod}{\text{ } \mathrm{Mod} \text{ }}

\begin{document}
\begin{frontmatter}

\title{
Quantization and Security Parameter Design for Overflow-Free Confidential FRIT 
\thanksref{footnoteinfo}
} 

\thanks[footnoteinfo]{This work was supported by JSPS KAKENHI Grant Numbers 22H01509 and 23K22779.}

\author[First]{Jugjin Park} 
\author[Second]{Osamu Kaneko} 
\author[Second]{Kiminao Kogiso}

\address[First]{Cluster II (Emerging Multi-Interdisciplinary Engineering),\\ School of Informatics and
Engineering, The University of Electro-Communications, Chofu, Tokyo, Japan\\ (e-mail: parkjungjin@uec.ac.jp).}
\address[Second]{Department of Mechanical and Intelligent Systems Engineering,\\ Graduate School of
Informatics and Engineering, The University of Electro-Communications, Chofu, Tokyo,
Japan\\ (e-mail: \{o.kaneko, kogiso\}@uec.ac.jp)}

\begin{abstract}
This study proposes a systematic design procedure for determining the quantization gain and the security parameter in the Confidential Fictitious Reference Iterative Tuning (CFRIT), enabling overflow-free and accuracy-guaranteed encrypted controller tuning.
Within an encrypted data-driven gain tuning, the range of quantization errors induced during the encoding (encryption) process can be estimated from operational data.
Based on this insight, explicit analytical conditions on the quantization gain and the security parameter are derived to prevent overflow in computing over encrypted data.
Furthermore, the analysis reveals a quantitative relationship between quantization-induced errors and the deviation between the gains obtained by CFRIT and non-confidential Fictitious Reference Iterative Tuning (FRIT), clarifying how parameter choice affects tuning accuracy.
A numerical example verifies the proposed procedure by demonstrating that the designed parameters achieve accurate encrypted tuning within a prescribed tolerance while preventing overflow.
In addition, the admissible region of parameter combinations is visualized to examine the characteristics of feasible and infeasible regions, providing practical insights into parameter design for encrypted data-driven control.
\end{abstract}


\end{frontmatter}

\section{Introduction}

%
A Cyber-Physical System (CPS) is a key enabler of Industry 5.0, where such systems are expected to operate not only autonomously but also adaptively, reflecting human intentions and responding to environmental variations in real time.
By leveraging CPS technologies, applications such as cooperative robotic systems working alongside humans~\citep{CoRobot2019}, industrial energy optimization with digital twins~\citep{twindigital2021}, and smart city management~\citep{smartcity23} are envisioned to become increasingly feasible.
These applications are supported by several enabling technologies, including edge/cloud computing, digital twins, the Internet of Everything (IoE), and big data analytics~\citep{Industry5.0survey22}, which facilitate continuous collection, processing, and analysis of operational data.

Data-driven control~\citep{coulson2019data,berberich2020data} has emerged as a promising paradigm for designing and updating controllers directly from operational data.
Representative approaches include Fictitious Reference Iterative Tuning (FRIT)~\citep{Kaneko13} and Virtual Reference Feedback Tuning (VRFT)~\citep{campi2002virtual}, which demonstrate the potential of data-driven tuning and are increasingly adopted in distributed environments.
While edge-side execution offers advantages in terms of low latency and reduced communication requirements, cloud-side computation is equally appealing because CPSs typically aggregate measurement and log data in cloud infrastructures.
Such integration of human-centered goals, adaptive control design, and cloud–edge data utilization aligns with the fundamental vision of Industry 5.0 systems that are tuned from data and evolve alongside human intentions.

Despite the growing success of data-driven control, its integration into CPSs has raised serious concerns about security and confidentiality~\citep{sandberg2015cyberphysical}.
Conventional countermeasures, such as encrypting communication channels~\citep{pang2011secure} and introducing redundancy-based mechanisms, are often insufficient to address CPS-specific threats.
In particular, poisoning attacks intentionally manipulate operational data to degrade control performance~\citep{fan2022survey,ikezaki2023poisoning}, posing a serious risk to data-driven approaches in which controllers are directly identified or adapted from input–output data.
These vulnerabilities highlight the need for security frameworks tailored to the specific characteristics of CPSs.
Among these, \cite{Hos25} proposed the Confidential FRIT (CFRIT) as a secure extension of FRIT, enabling controller tuning directly over encrypted data. 

Although the CFRIT framework successfully ensures data confidentiality, they do not consider the computational reliability of encrypted/encoded data processing, particularly potential overflow and precision loss caused by quantization and the security parameter. 
In CFRIT, tuning performance is highly sensitive to these parameters, and inappropriate choices of the quantization gain or the security parameter can severely degrade computational accuracy, resulting in overflow and significant deviation from the ideal gain. 
However, \cite{Hos25} did not discuss how to systematically design these parameters to achieve accurate and overflow-free tuning. 
Establishing such a design procedure is essential to reduce trial-and-error efforts and to ensure secure and reliable encrypted computations.

The objective of this study is to propose a systematic design procedure for determining the quantization gain and the security parameter in CFRIT, ensuring overflow-free and accuracy-guaranteed controller tuning.
The key idea is that, within a data-driven control framework, the range of quantization errors induced during the encoding (encryption) process can be estimated from operational data.
Based on this insight, explicit conditions on the quantization gain and the security parameter are derived to prevent overflow during encrypted computation.
Furthermore, the analysis clarifies how quantization errors affect the deviation between the gains obtained by CFRIT and traditional FRIT, revealing an analytical relationship between quantization error and tuning accuracy.
These results enable CFRIT to be systematically configured as a reliable encrypted tuning framework with guaranteed accuracy, ensuring that the tuned gain remains within a prescribed tolerance of that obtained by the traditional FRIT.
A numerical example verifies the proposed procedure by demonstrating that the resulting parameters achieve a tuning result within a prescribed tolerance under encryption while preventing overflow.
In addition, the example is used to visualize the admissible region of parameter combinations and to discuss the characteristics of the feasible and infeasible regions, which provide insights for future investigation.

The contributions of this study are twofold:
(i) We extend the concept of overflow to the context of encrypted data-driven control and derive explicit analytical conditions on the quantization gain and the security parameter that guarantee overflow-free computation.
Furthermore, we clarify the quantitative relationship between quantization-induced errors and the deviation between CFRIT and traditional FRIT tuning, thereby providing practical design guidelines for parameter selection.
(ii) Based on these analytical results, we propose a systematic parameter design procedure for CFRIT and verify its effectiveness through numerical examples, demonstrating accuracy-guaranteed encrypted controller tuning without overflow and visualizing the characteristics of the admissible parameter region.




The structure of this paper is as follows.
\textrm{Section~\ref{sec:pre}} provides the notations and  ElGamal encryption as preliminaries.
\textrm{Section~\ref{sec:prob}} provides a brief overview of FRIT and CFRIT, and formulates the parameter design problem for CFRIT.
\textrm{Section~\ref{sec:analysis}} proposes the systematic parameter design procedure based on the analysis result.
\textrm{Section~\ref{sec:exp}} verifies the proposed procedure and discusses the admissible parameter regions using a numerical example.
\textrm{Section~\ref{sec:con}} concludes this paper.

\section{Preminaries}\label{sec:pre}
\subsection{Notations}
Sets of real numbers, integers, plaintext spaces, and ciphertext spaces are denoted by $\mathbb{R},\mathbb{Z},\mathcal{M},\mathcal{C}$, respectively.
We define $\mathbb{R}^+ \coloneqq \{x \in \mathbb{R} \; | \; 0 < x\}$, 
$\mathbb{Z}^+ \coloneqq \{z \in \mathbb{Z} \; | \; 0 < z\}$, 
$\mathbb{Z}_n \coloneqq \{z \in \mathbb{Z} \; | \; 0 \leq z < n\}$, 
$\mathbb{Z}_n^+ \coloneqq \{z \in \mathbb{Z} \; | \; 0 < z < n\}$, and 
$\mathfrak{B}_a^b \coloneqq \{a^i \bmod b \; | \; i \in \mathbb{Z}_b\}$.
For a scalar $a \in \mathbb{R}$, its absolute value is denoted by $|a|$.
A set of vectors of size $n$ is denoted by $\mathbb{R}^n$. 
The $j$th element of vector $v$ is denoted by $v_j$.
$\ell_1$ norm, $\ell_2$ norm and max norm $v$ are denoted by $\|v\|_1$, $\|v\|_2$ and $\|v\|_{\max}$.
The set of matrices of size $m \times n$ is denoted by $\mathbb{R}^{m \times n}$.
An $(i,j)$-entry of matirx $M$ is denoted by $M_{ij}$.
The maximum norm of $M$ is denoted by $\|M\|_{\max} \coloneqq \max_{(i,j)}(|M_{ij}|)$, respectively, and the minimum eigenvalue of $M$ is denoted by $\lambda_{\min}(M)$.
For a permutatiton $\sigma=\begin{pmatrix}
    1 & 2 & \cdots & n \\
    i_1 & i_2 & \cdots & i_n
\end{pmatrix}$, $\mathrm{sgn}\sigma=1$ if $\sigma$ is an even permutation; otherwise, $-1$ returns.
The determinant of a matrix $X \in \mathbb{R}^{n \times n}$ is expressed as $\det(X) = \sum_{\sigma \in \mathbb{S}^n}\{(\mathrm{sgn} \sigma)\prod_{i=1}^n x_{i\sigma(i)}\}$, where $\sigma$ represents a permutation and $\mathbb{S}^n$ is the symmetric group of degree $n$ (the automorphism group of $\{1,\cdots,n\}$).
The greatest common divisor of the two positive integers $a,b \in \mathbb{Z}^+$ is denoted by $\gcd(a,b)$.

The minimal residue of integer $a \in \mathbb{Z}$ modulo $m \in \mathbb{Z}^+$ is defined as
\begin{equation*}
    a \bMod m \coloneqq 
    \begin{cases}
        b \quad & \mathrm{if} \quad  b < |b-m|, \\
        b -m \quad & \mathrm{otherwise,}
    \end{cases}
\end{equation*}
where $b = a \bmod m$. 

    Let $p$ be an odd prime number and $z$ be an integer satisfying $\gcd(z,p) = 1$.
    If there exists an integer $b$ such that $b^2 = z \bmod p$, then the integer $z$ is a quadratic residue of modulo $p$.
    This can be expressed using the Legendre symbol $(\cdot /\cdot)_L$ as follows:
    \begin{align*}
        \bigg( \frac{z}{p} \bigg)_L &\coloneqq z^{\frac{p-1}{2}} \bMod p \\
        &= \begin{cases}
            1 & \quad \text{if $z$ is a quadratic residue,} \\
            -1 & \quad \text{if $z$ is a quadratic nonresidue.}
        \end{cases}
    \end{align*}

    The rounding function $\left \lceil \cdot \right \rfloor$ of $\sigma \in \mathbb{R}^+$ to the nearest positive integer is defined as
    \begin{equation}
        \left \lceil \sigma \right \rfloor =
        \begin{cases}
            \left \lfloor \sigma + 0.5 \right  \rfloor & \quad \text{if $\sigma \geq 0.5$,} \\
            1 & \quad \text{otherwise,}
        \end{cases}
        \label{eq:round}
    \end{equation}
    where $\left \lfloor \cdot \right \rfloor$ denotes the floor function.

\subsection{Quantizer}
\begin{figure}
\begin{center}
\includegraphics[width=8.4cm]{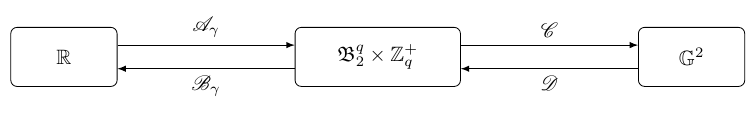}
\caption{Relationship of the maps between the real and cyclic groups} 
\label{fig:quantization}
\end{center}
\end{figure}

A quantizer is required to construct the encryption control system because the plaintexts and ciphertexts in the encryption scheme are integers, and the processes at the controller are reconstructed using the encryption scheme. 
Inspired from \cite{quantization01}, we define encoding maps $\mathscr{A}_\gamma$ and $\mathscr{C}$, and decoding maps $\mathscr{B}_\gamma$ and $\mathscr{D}$ as
\begin{align*}
   \mathscr{A}_\gamma & : \mathbb{R} \rightarrow \mathfrak{B}_2^q \times \mathbb{Z}_q^+, \\
   & : x \mapsto \begin{cases}
      (1, \left \lceil \gamma |x| \right \rfloor) \quad & \text{if } x \geq 0, \\
      (2, \left \lceil \gamma |x| \right \rfloor) \quad & \text{if } x < 0, 
   \end{cases} \\
   \mathscr{B}_\gamma & : \mathfrak{B}_2^q \times \mathbb{Z}_q^+ \rightarrow \mathbb{R}, \\
   & : (\zeta , z) \mapsto \bigg(\frac{\zeta}{3} \bigg)_L \frac{z}{\gamma} \coloneqq \breve{x} , \\
   \mathscr{C} & : \mathfrak{B}_2^q \times \mathbb{Z}_q^+ \rightarrow \mathbb{G}^2 ,\\
   & : (\zeta, z) \mapsto \bigg(\bigg(\frac{\zeta}{p} \bigg)_L \zeta, \bigg(\frac{z}{p}\bigg)_L z \bigg) \bmod p \coloneqq (\bar{x}_1 , \bar{x}_2), \\
   \mathscr{D} & : \mathbb{G}^2 \rightarrow \mathfrak{B}_2^q \times \mathbb{Z}_q^+ , \\
   & : (\bar{x}_1, \bar{x}_2) \mapsto (|\bar{x}_1 \bMod p|, |\bar{x}_2 \bMod p|),
 \end{align*}
where $\gamma \geq 1$ is the quantization gain, $\zeta \in \{1,2\}$, and $z \coloneqq \left \lceil \gamma |x| \right \rfloor \bmod q$.
we employ $\Ecd \coloneqq \mathscr{C} \circ \mathscr{A}_\gamma$ and $\Dcd \coloneqq \mathscr{B}_\gamma \circ \mathscr{D}$,
which perform elementwise operations for a vector and a matrix.
The relationship between the maps is shown in Fig. \ref{fig:quantization}.

\subsection{ElGamal Encryption}
The ElGamal encryption scheme, proposed in~\cite{elgamal1985public}, is public-key multiplicatively homomorphic encryption and defined as a tuple $\mathcal{E}  \coloneqq (\KeyGen,\Enc,\Dec)$,
where 
$\KeyGen:\mathcal{S}\rightarrow\mathcal{K}=\mathcal{K}_\mathsf{p}\times\mathcal{K}_\mathsf{s}:k\mapsto(\pk,\sk)=((p,q,g,h),s)$ is a key generation algorithm, 
$\Enc:\mathcal{K}_p\times\mathcal{M}\rightarrow\mathcal{C}:(\pk,(\bar{x}_1,\bar{x}_2))\mapsto c=((g^{r_1}\bmod p,\bar{x}_1 h^{r_1}\bmod p),(g^{r_2}\bmod p,\bar{x}_2 h^{r_2}\bmod p))$ is encryption algorithm, 
$\Dec:\mathcal{K}_s\times\mathcal{C}\rightarrow\mathcal{M}:(\sk,(c_1,c_2),(c_3,c_4))\mapsto(c_1^{-s}c_2\bmod p,\;c_3^{-s}c_4\bmod p)$ is a decryption algorithm, 
$\pk$ is a public key, 
$\sk$ is a secret key, 
$q$ is a $\kappa$-bit prime\footnote{From a security standpoint, $q$ is regarded as the largest prime representable within $\kappa$-bit integer variables.
In a numerical example, we evaluate $\kappa$ instead of $q$.}, 
$p=2q+1$ is a safe prime, 
$g$ is generator of a cyclic group $\mathbb{G}\coloneqq\{g^i\bmod p\; | \; i\in\mathbb{Z}_q\}$ such that $g^q\bmod p=1$, $h=g^s\bmod p$, 
$\mathcal{M}=\mathbb{G}^2$,
$\mathcal{C}=\mathbb{G}^4$, and 
$r_1$, $r_2$, and $s$ are random numbers chosen from $\mathbb{Z}_q$.
$\Enc$ and $\Dec$ perform elementwise operations for a vector and matrix.
Note that for $m$, $m^{\prime}\in\mathcal{M}$, the ElGamal encryption possesses multiplicative homomorphism:
$\Dec(\sk,\Enc(\pk,(\bar{x}_1,\bar{x}_2))*\Enc(\pk,(\bar{x}_1^{\prime},\bar{x}_2^{\prime})\bmod p)=(\bar{x}_1 \bar{x}_1^{\prime}, \bar{x}_2 \bar{x}_2^{\prime})\bmod p$, where $*$ denotes the Hadamard product.
Hereafter, for simplicity, the arguments $\pk$ and $\sk$ of $\Enc$ and $\Dec$ will be omitted.

\section{Parameter Design Problem for CFRIT}\label{sec:prob}
This section formulates the parameter design problem for CFRIT, after briefly introducing the essentials of both the traditional (non-confidential) and the confidential FRIT methods.

\subsection{Problem Setup for Gain Tuning}
We consider a linear plant in the discrete-time state-space representation:
\begin{align}
x(t+1)=Ax(t)+Bu(t), \label{eq:plant}
\end{align}
where $t\in\mathbb{Z}^+$, $x\in\mathbb{R}^n$, and $u\in\mathbb{R}$ denote the step, state (measurable), and control input, respectively. 
It is assumed that the initial state is zero, i.e., $x(0)=0$, and that the system order of a minimal realization is known, while the plant model \eqref{eq:plant} is unknown.
For the plant \eqref{eq:plant}, the following controller structure is designed to achieve $\lim_{t\rightarrow\infty}x(t)=0$,
\begin{align}
u(t)=Fx(t)+v(t), \label{eq:controller}
\end{align}
where $F\in\mathbb{R}^{1 \times n}$ is the state-feedback gain, and $v\in\mathbb{R}$ is an intentionally applied signal used by the designer to evaluate the closed-loop system.
The closed-loop transfer function is given by $H(F)\coloneqq\tfrac{x(z)}{v(z)}=(zI-A-BF)^{-1}B\in\mathbb{R}^n[z]$.
The control system designer is assumed to have access to the system and its data $u$ and $x$, as well as an initial state-feedback gain $F_{\mathrm{ini}}$.
The designer aims to obtain a new gain $F^*$ that achieves the desired closed-loop characteristics $H^*$.

Let $x(t;F)$ and $u(t;F)$ denote the state and input at step $t\in\mathbb{Z}^+$ associated with the feedback gain $F$, respectively. 
The control system designer constructs the vector $E$ and matrix $W$ from $N$ samples of $n$-dimension signals of $x(t;F_{\mathrm{ini}})$ and $u(t;F_{\mathrm{ini}})$ as follows:
\begin{align*}    
E&\coloneqq
\begin{bmatrix}
    e_1 \\ e_2 \\ \vdots \\ e_n
\end{bmatrix} \in \mathbb{R}^{nN}, \quad 
W\coloneqq 
\begin{bmatrix}
    w_1 \\ w_2 \\ \vdots \\ w_n
\end{bmatrix} \in \mathbb{R}^{nN \times n}, 
\end{align*}
where
\begin{align*}
  e_j & = 
  \begin{bmatrix}
    x_j(t;F_\mathrm{ini})-H_{\mathrm{d}j}u(t;F_\mathrm{ini})\\
    \vdots\\
    x_j(t+N;F_\mathrm{ini})-H_{\mathrm{d}j}u(t+N;F_\mathrm{ini})
  \end{bmatrix}
  \in\mathbb{R}^{N},\\
w_j & = 
  \begin{bmatrix}
    H^*_jx(t;F_\mathrm{ini})^\top\\
    \vdots\\
    H^*_jx(t+N;F_\mathrm{ini})^\top
  \end{bmatrix}
  \in\mathbb{R}^{N\times n},\quad \forall j\in\{1,\cdots,n\}.
\end{align*}
The vector $E$ represents the difference between the initial state and the target state, 
and the matrix $W$ represents the response of $H^*$ to the applied initial state $x_{\mathrm{ini}}$.

\subsection{Traditional FRIT}
The traditional FRIT method, proposed in~\cite{Soma04}, determines a new feedback gain $F^*$ that minimizes the objective function $J(F)=\|x(t;F_{\mathrm{ini}})-H^*v(t;F)\|_2=\|(H(F)-H^*)v(t;F)\|_2$, where the pseudo-exogenous signal is defined as $v(t;F)=u(t;F_{\mathrm{ini}})-Fx(t;F_{\mathrm{ini}})$.
Under the mild assumption that $v$ contains sufficient excitation modes, the minimizer $F^*$ of $J(F)$ yields a closed-loop system $H(F^*)$ that sufficiently approximates $H^*$.
Based on the least-squares method, the minimizer $F^*$ of $J(F^*)=(E+WF^{*^{\top}})^\top(E+WF^{*^{\top}})$ is obtained as 
\begin{align}
 \label{eq:feedback_gain_FRIT}
  F^*=-E^\top W(W^\top W)^{-1}=-E^\top W\Phi.
\end{align}
where $\Phi:=(W^\top W)^{-1}$.

\subsection{Confidential FRIT}
The CFRIT method, proposed in~\cite{Hos25}, determines a new gain $F_\mathcal{E}^*$ based on the computation over encrypted data using the multiplicative homomorphism of the ElGamal encryption scheme.
However, the computation in \eqref{eq:feedback_gain_FRIT} poses difficulties when executed over encrypted data 
due to the inverse matrix $\Phi=(W^\top W)^{-1}$.
To address this issue, an alternative scalar representation of the inverse matrix is derived by introducing intermediate generation matrices $\Phi_k\in\mathbb{R}^{n\times n}$ satisfying $(W^\top W)^{-1}=\sum_{k=1}^{(n-1)!}\Phi_{k}$.
Using cofactor expansion and determinant computation, the $(j,i)$-element of $\Phi_k$ is given by
\begin{align*}
\Phi_{k,ji}=\det(\Psi)^{-1}(-1)^{i+j} \mathrm{sgn}(\sigma_k)\prod_{\xi=1}^{n-1} \tilde{\Psi}_{\xi,\sigma_k(\xi)},
\end{align*}
where $\mathrm{sgn}(\sigma_k)$ denotes the sign of the permutation $\sigma_k$ corresponding to the $k$-th row of the signed permutation matrix, and $\tilde{\Psi}_{\xi,ij}$ is the determinant obtained by removing the $i$-th row and $j$-th column of $\Psi:= W^\top W$.

To decompose the computation of $F^*$ \eqref{eq:feedback_gain_FRIT} into addition and multiplication, the $\iota$-th element of $F^*$, denoted as $F^*_\iota$, is considered:
\begin{align*}
F_\iota^* 
&=-\sum_{k=1}^{(n-1)!}\sum_{i=1}^{nN} \sum_{l=1}^{n}E_i W_{il}\Phi_{k,l\iota},\\
&=-\sum_{k=1}^{(n-1)!}\sum_{i=1}^{nN} \sum_{l=1}^{n}E_i W_{il} \det(\Psi)^{-1}\times\\
&\hspace*{19ex}(-1)^{l+\iota}\mathrm{sgn}(\sigma_k)\prod_{\xi=1}^{n-1} \tilde{\Psi}_{\xi,\sigma_k(\xi)},\\
&=\sum_{j=1}^{M} \mathsf{F}_{j\iota},
\end{align*}
where each $\mathsf{F}_{j\iota}\in\mathbb{R}$ is the multiplicative term define as
\begin{align}
\mathsf{F}_{j\iota}\coloneqq
-E_iW_{il}\det(\Psi)^{-1}(-1)^{l+\iota} \mathrm{sgn}(\sigma_k)\prod_{\xi=1}^{n-1} \tilde{\Psi}_{\xi,\sigma_k(\xi)},
\label{eq:F_jl}
\end{align}
and $j$ is the index up to $M=(n-1)!n^2N$ representing the triplet $j(k,i,l)=(k-1)nN+(i-1)n+l$, $\forall k\in\{1,\cdots,(n-1)!\}$, $\forall i\in\{1,\cdots,nN\}$, and $\forall l\in\{1,\cdots n\}$.

The CFRIT method enables computation of the multiplication term \eqref{eq:F_jl} over encoded/encrypted data, and thus \textit{Proposition~4.2} in \cite{Hos25} leads to the following homomorphic operation:
$\Enc(\Ecd(\mathsf{F}_{j\iota}))\coloneqq\Enc(\Ecd(E_i))*\Enc(\Ecd(W_{il}))*\Enc(\Ecd(\det(\Psi)^{-1}))*\Enc($\\$\Ecd((-1)^{l+\iota}))*\Enc(\Ecd(\mathrm{sgn}(\sigma_k)))*\bigotimes_{\xi=1}^{n-1}\tilde{\Psi}_{\xi,\sigma_k(\xi)}$.
This results in the tuned gain $F_\mathcal{E}^*$ obtained via the homomorphic operations:
\begin{align}
\label{eq:feedback_gain_CFRIT}
F_{\mathcal{E}}^* =
\begin{bmatrix}
  \sum_{j=1}^{M} \Dcd(\Dec(\Enc(\Ecd(\mathsf{F}_{j1})))) \\
    \vdots \\
 \sum_{j=1}^{M} \Dcd(\Dec(\Enc(\Ecd(\mathsf{F}_{jn})))) \\
\end{bmatrix}^{\!\top},
\end{align}
where addition is conducted after decryption and decoding.
In this study, the multiplicative term \eqref{eq:F_jl} will be used to analyze overflow avoidance in Section~\ref{sec:overflow}.

\begin{figure}
    \centering
    \begin{minipage}{0.5\textwidth}
        \centering
        \includegraphics[width=\textwidth]{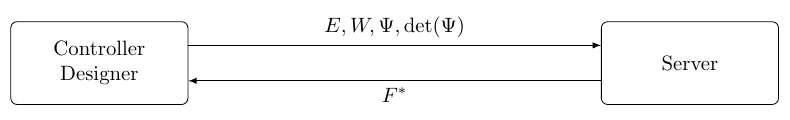}
        \subcaption{Traditional FRIT}
        \label{fig:FRIT}
    \end{minipage}
    \begin{minipage}{0.5\textwidth}
        \centering\vspace*{2ex}
        \includegraphics[width=\textwidth]{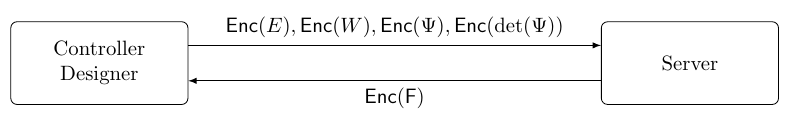}
        \subcaption{CFRIT}
        \label{fig:CFRIT}
    \end{minipage}
    \caption{Configurations of FRIT and CFRIT}
    \label{fig:FRIT_comparison}
\end{figure}

\subsection{Parameter Design Problem}
An inappropriate choice of the parameters may lead to overflow.
This study addresses the design of the quantization gain $\gamma$ and the security parameter $\kappa$ such that the deviation between the gain obtained by the traditional FRIT and that obtained by the CFRIT remains within a prescribed bound, while preventing overflow. 
\begin{prob}\label{prob}
Let $F^*$ denote the state-feedback gain obtained from FRIT, and let $F_{\mathcal{E}}^*(\kappa,\gamma)$ denote the state-feedback gain obtained from CFRIT using the ElGamal encryption scheme $\mathcal{E}$ with security parameter $\kappa$ and encryption/decryption operations $\Enc$ and $\Dec$ under quantization gain $\gamma$.  
Given datasets $E$ and $W$ and a prescribed tolerance $\epsilon>0$, the problem is to determine $\kappa$ and $\gamma$ such that
$\|F_{\mathcal{E}}^*(\kappa,\gamma)-F^*\|_2\leq\epsilon$
is satisfied without overflow.
\end{prob}
This study aims to provide a systematic parameter design procedure to solve~\textit{Problem~\ref{prob}}, based on the quantization error analysis presented in the following section. 
In addition, \cite{Hos25} provides no discussion regarding the choice of $\kappa$ and $\gamma$. 

\section{Parameter Design Procedure}\label{sec:analysis}
This section clarifies the relationship among the parameters $\kappa$, $\gamma$, and the prescribed tolerance $\epsilon$, derives the conditions required to prevent overflow, and then proposes the parameter design procedure.

\subsection{Overflow-Avoidance Conditions}\label{sec:overflow}
We derive the conditions on the quantization gain $\gamma$ and the security parameter $\kappa$ that guarantee overflow prevention.
Consider the ElGamal encryption scheme with modulus $q$, which is a $\kappa$-bit prime. 
The overflow is defined as follows.
\begin{defn}\label{def:multioverflow}
An overflow occurs if, for real values $x_i\in\mathbb{R}$, $\forall i\in\{1,\cdots,n+1\}$ and a quantization gain $\gamma\in\mathbb{Z}^+$,   
the inequality $\prod_i^{n+1}\left\lceil\gamma |x_i|\right\rfloor\geq q-\tfrac{1}{2}$ holds.
\end{defn}
In this definition, we use the fact that the rounding function~\eqref{eq:round} implies that $\lceil\sigma\rfloor \geq 1$ for any positive $\sigma$. 
A sufficient condition for choosing the parameters to avoid overflow, which is the fundamental result of this study, is given as follows.
\begin{thm}\label{thm:gammachoice}
No overflow occurs if $(q,\gamma)\in Q$ holds, where 
\begin{align}\label{ieq:security_parameter_Design}
Q:=\left\{(q,\gamma)\,\middle|\,q- \left\lceil\gamma^{n+5}\frac{\|E\|_{\max}\,\|W\|_{\max}}{\lambda_{\min}(\Psi)}\right\rfloor>\frac{1}{2}\right\}.
\end{align}
\end{thm}

\begin{pf}
From \textit{Definition~\ref{def:multioverflow}}, the condition of occurring no overflow is given by
$\prod_i^{n+1}\left\lceil\gamma |x_i|\right\rfloor < q-\tfrac{1}{2}.$
$\mathsf{F}_{j\iota}$ is computed via multiplicative homomorphism, and so $x_i$ is considered as each term of the right-hand side of \eqref{eq:F_jl}, such as $x_1:=-1$, $x_2:=E_{i}$, $x_3 \coloneqq W_{il}$.
The inequality condition is obtained as follows:
\begin{align}
\left\lceil\gamma^{n+5} |\mathsf{F}_{j\iota}|\right\rfloor < q -\tfrac{1}{2},\label{noc}
\end{align} 
where we use the approximation
\begin{align}
\prod_{i=1}^{n+1}\left\lceil\gamma |x_i|\right\rfloor\approx \left\lceil\prod_{i=1}^{n+1}\gamma |x_i|\right\rfloor, \nonumber
\end{align}
for the sufficiently large quantization gain $\gamma$.
Subsequently, we confirm that $|\mathsf{F}_{j\iota}|$ is bounded:
\begin{align*}
   |\mathsf{F}_{j\iota}| &= \big|E_i W_{il} \det(\Psi)^{-1} (-1)^{l+\iota} \mathrm{sgn}(\sigma_k)
    \prod_{\xi=1}^{n-1} \tilde{\Psi}_{\xi,\sigma_k(\xi)} \big| \\
   &= |E_{i}| \, |W_{il}| \, \big|\det(\Psi)^{-1} \mathrm{sgn}(\sigma_k) \prod_{\xi=1}^{n-1} \tilde{\Psi}_{\xi, \sigma_{k}(\xi)} \big|, \\
   &= |E_{i}| \, |W_{il}| \, |\Phi_{k,l\iota}|, \\
   &\leq \|E\|_{\max} \, \|W\|_{\max}\, \|\Phi\|_{\max}, \\
   &\leq\frac{\|E\|_{\max}\, \|W\|_{\max}}{\lambda_{\min}(\Psi)}, 
\end{align*}
where we used the following condition \citep{Matrix01}: 
$|\Phi_{k,l\iota}|\leq \|\Phi\|_{\max} \leq \lambda^{-1}_{\min}(\Psi)$, $\forall k,l,\iota$.
Hence, \eqref{noc} can be written into the sufficient condition \eqref{ieq:security_parameter_Design} 
to prevent overflow.
\hfill $\qed$
\end{pf}

\begin{rem}\label{rem:flexibility}
The parameter set $Q$ in~\eqref{ieq:security_parameter_Design} implies flexibility in choosing $(q,\gamma)$ to prevent overflow. 
By introducing an objective function such as computational cost, the parameter design problem can be extended to determine the optimal combination.
This extension requires further discussion on the definition of the objective function and the development of an efficient search algorithm, which is left as future work.
\end{rem}

\subsection{Quantization Error Analysis}\label{sec:design_CFRIT} 
To design the quantization gain according to a prescribed error tolerance, it is necessary to estimate the quantization error. 
We first derive bounds on the quantization error.
\begin{lem}\label{lem:5}
If $\breve{x}x\geq 0$, then $\bigl|\,|\breve{x}|-|x|\,\bigr|=|\breve{x}-x|$.
\end{lem}
\begin{pf}
Since $\breve{x}x\geq 0$, we have $|\breve{x}x|=\breve{x}x$.  
Expanding the square gives
$(|\breve{x}|-|x|)^2=|\breve{x}|^2-2|\breve{x}||x|+|x|^2=\breve{x}^2-2\breve{x}x+x^2=(\breve{x}-x)^2$.
Taking square roots on both sides yields
$\bigl|\,|\breve{x}|-|x|\,\bigr|=|\breve{x}-x|$.
\hfill $\qed$
\end{pf}

\begin{lem}\label{lem:quantization error scalar}
Let the quantization error be $\Delta\coloneqq|\breve{x}-x|$.
If $x\in\mathbb{R}$ satisfies $\breve{x}x\geq 0$ under the quantizer, then
\begin{align}
0\leq\Delta\leq\frac{1}{\gamma}.\label{lem:qes}
\end{align}
\end{lem}
\begin{pf}
By \textit{Lemma~\ref{lem:5}}, we have $\Delta=\bigl|\,|\breve{x}|-|x|\,\bigr|$ when $\breve{x}x\geq 0$.
The quantizer is defined as $\breve{x}=\left(\frac{\zeta}{3}\right)_L\frac{z}{\gamma}$ with $z=\left\lceil\gamma |x|\right\rfloor\bmod q$.
Assuming no overflow, this reduces to $\Delta=\bigg|\frac{\left\lceil\gamma |x|\right\rfloor}{\gamma}-|x|\bigg|$.
\textbf{Case 1:} $\gamma |x|\geq\tfrac{1}{2}$.  
From the definition of rounding, $\left\lceil\gamma |x|\right\rfloor=\lfloor\gamma |x|+0.5\rfloor$, so $0\leq\Delta\leq\frac{1}{2\gamma}$.
\textbf{Case 2:} $\gamma |x|<\tfrac{1}{2}$.  
In this case $\left\lceil\gamma |x|\right\rfloor=1$, and hence
$\Delta=\frac{1-\gamma |x|}{\gamma}$, $\frac{1}{2\gamma}<\Delta\leq\frac{1}{\gamma}$.
Combining both cases yields~\eqref{lem:qes}.
\hfill $\qed$
\end{pf}

\begin{rem} 
In~\cite{teranishi2019}, the maximum width of $\mathcal{M}$ is set to $d_{\max}=1$, which leads to inequality~\eqref{lem:qes}.
In that quantizer, zero is mapped to zero so that $\lceil \gamma |x|\rfloor=0$ holds when $\gamma|x|<0.5$.
However, in the ElGamal encryption scheme, encrypting zero yields a ciphertext of zero, thereby revealing the plaintext and posing a potential security risk.
To address this issue, the CFRIT adopts a modified rounding quantizer~\citep{quantization01} that prohibits zero output.
Specifically, when $\gamma|x|<0.5$, the quantized value is defined as $\lceil\gamma |x|\rfloor=1$ instead of 0.
Consequently, an input of zero, or a small input that would otherwise be quantized to zero due to an insufficient quantization gain, is converted into a nontrivial ciphertext, thereby preserving the desired security property.
\end{rem}

We next derive the relationship between the quantization error $\Delta$ and $\|F_{\mathcal{E}}^*(\kappa,\gamma)-F^*\|_2$, summarized in the following theorem.
\begin{thm}
Assume that $\kappa$ is set to prevent overflows.
If, for given $\epsilon$, the quantization gain $\gamma$ is chosen from the following set~$\Gamma(\epsilon)$:
\begin{align}\label{ieq:quantization_gain_Design}
\Gamma(\epsilon):=\bigg\{\gamma\geq 1\,\bigg|\,\gamma\geq\frac{Mn}{\epsilon}\bigg\},
\end{align}
then $\|F_{\mathcal{E}}^*(\kappa,\gamma)-F^* \|_2 \leq\epsilon$ holds.
\end{thm}
\begin{pf}
The assumption that $\kappa$ prevents overflows ensures that $F^*_\mathcal{E}$ can be computed correctly under the ElGamal encryption scheme.
From \eqref{eq:feedback_gain_CFRIT}, 
\begin{equation*}
    F_{\mathcal{E}}^* =
   \begin{bmatrix}
      \sum_{j=1}^{M} \breve{\mathsf{F}}_{j1}  &
      \sum_{j=1}^{M} \breve{\mathsf{F}}_{j2}  &
      \cdots &
      \sum_{j=1}^{M} \breve{\mathsf{F}}_{jn}  
   \end{bmatrix}.
\end{equation*}
Hence, the difference between $F_{\mathcal{E}}^*$ and $F^*$ is given by
\begin{equation*}
    F_{\mathcal{E}}^* - F^* =
   \begin{bmatrix}
       \sum_{j=1}^{M}(\breve{\mathsf{F}}_{j1} - \mathsf{F}_{j1}) &\cdots&
       \sum_{j=1}^{M}(\breve{\mathsf{F}}_{jn} - \mathsf{F}_{jn})
   \end{bmatrix}.
\end{equation*}
From the definition of quantization error, 
$\Delta_{j\iota}=|\breve{\mathsf{F}}_{j\iota}-\mathsf{F}_{j\iota}|$, and  \textit{Lemma~\ref{lem:quantization error scalar}} gives 
$0\leq\Delta_{j\iota}\leq\tfrac{1}{\gamma}$.
Therefore,
\begin{align*}
||F_{\mathcal{E}}^*-F^*||_1=\sum_{\iota=1}^n\sum_{j=1}^{M}\Delta_{j\iota}\leq\frac{Mn}{\gamma},
\end{align*}
Since $\|\nu\|_2\leq\|\nu\|_1$ for any vector $\nu$, we have $\|F_{\mathcal{E}}^*-F^*\|_2\leq\tfrac{Mn}{\gamma}$. 
Thus, choosing $\gamma$ to satisfy $\gamma>\frac{Mn}{\epsilon}$ ensures $\|F_{\mathcal{E}}^*-F^*\|_2<\epsilon$.
\hfill$\qed$
\end{pf}

\subsection{Step-by-Step Design Procedure}\label{sec:proposed}
The proposed procedure provides a systematic approach for determining the quantization gain and security parameter to ensure both overflow-free computation and a prescribed error tolerance. 
The procedure is summarized as follows.

\begin{enumerate}
\item[\textsf{1.}] Consider a state-feedback control system with an initial feedback gain $F_{\mathrm{ini}}$.
\item[\textsf{2.}] Specify the desired closed-loop characteristics $H^*$ and the allowable error tolerance $\epsilon$. 
\item[\textsf{3.}] Design an external signal $v(t)$, $t \in \{0,1,\cdots,N-1\}$, that sufficiently excites the control system.
\item[\textsf{4.}] Apply $v(t)$ to the plant and measure the corresponding state and input signals.
\item[\textsf{5.}] Compute $E$ and $W$ using the measured signals.
\item[\textsf{6.}] Choose $\bar\gamma$ from the admissible set $\Gamma$ in~\eqref{ieq:quantization_gain_Design}.
\item[\textsf{7.}] For $\bar\gamma$, choose $\bar\kappa$ from the admissible set $Q$ in~\eqref{ieq:security_parameter_Design}.
\item[\textsf{8.}] Execute CFRIT with $\bar\kappa$ and $\bar\gamma$ to obtain $F_\mathcal{E}^*$.
\end{enumerate}

The proposed procedure enables the determination of $F_\mathcal{E}^*$, which theoretically guarantees the desired closed-loop performance within the specified tolerance $\epsilon$, without explicitly computing the original gain $F^*$ obtained by the traditional FRIT method.

\section{Numerical Example}\label{sec:exp}
This section presents a numerical example to demonstrate the proposed design procedure and to visualize an admissible parameter region.
The used computer was a MacBook Air equipped with an Apple M4 chip and 24 GB of memory, running macOS 15.6.
The CFRIT simulations were implemented using our developed C++ Encrypted Control Library (Version 3.3).

\subsection{Parameter Design}
The closed-loop system is given by
\begin{align*}
       A  &=  \begin{bmatrix}
         0.6018 & 0.4488 & 0.0926 & 0.1290 \\
         -0.5726 & 0.0291 & 0.1238 & 0.2164 \\
         0.0556 & 0.0774 & 0.9166 & 0.8875 \\
         0.0743 & 0.1298 & -0.1517 & 0.7649
      \end{bmatrix}, \ 
      B = \begin{bmatrix}
         0.1019 \\ 0.1496 \\ 0.0093 \\ 0.0258
      \end{bmatrix},\\
      F_\mathrm{ini}  &= \begin{bmatrix}     0.0283 & -0.0102 & -0.0023 & -0.0892 \end{bmatrix}.
\end{align*}
The initial conditions were set to $x(0)=0\in\mathbb{R}^4$ ($n=4$) and $v(0)=0$, with a sampling period of 1.0 $\mathrm{s}$.
The desirable closed-loop characteristics were specified as $H^*=\begin{bmatrix}H_{x_1}^* & H_{x_2}^* & H_{x_3}^* & H_{x_4}^*\end{bmatrix}^\top$, where
\begin{align*}
    H_{x_1}^* & = 
        \frac{0.1019z^3	-0.1029z^2	-0.02190z + 0.05256}{z^4	-2.3843z^3 + 2.2404z^2	-1.0107z + 0.2395}, \\
    H_{x_2}^* & = 
        \frac{0.1506z^3	-0.3932z^2 +	0.3647z	-0.1211}{z^4	-2.3843z^3 + 2.2404z^2	-1.0107z + 0.2395}, \\ 
    H_{x_3}^* & = 
        \frac{0.0093z^3 +	0.02719z^2	-0.0043z	-0.0026}{z^4	-2.3843z^3 + 2.2404z^2	-1.0107z + 0.2395}, \\ 
    H_{x_4}^* & = 
        \frac{0.0258z^3	-0.0143z^2	-0.0191z +	0.0077}{z^4	-2.3843z^3 + 2.2404z^2	-1.0107z + 0.2395}.
\end{align*}
The tolerance was set to $\epsilon=10^{-5}$, and the external input was defined as $v(t)=1$ for $1\leq t\leq 5$; and $v(t)=0$ otherwise.
The datasets $E$ and $W$ were constructed from the fifty samples ($N=50$) of $u$ and $x$, resulting in $M=4800$.
The above setup corresponds to the operations up to \textsf{Step~5} of the proposed procedure.
At \textsf{Step~6}, using~\eqref{ieq:quantization_gain_Design}, the admissible set was obtained as $\Gamma(\epsilon)=\{\gamma\,|\,\gamma\geq 1.92\times 10^9\}$, and $\bar{\gamma}=1.92\times 10^{9}$ was chosen.
At \textsf{Step~7}, using~\eqref{ieq:security_parameter_Design}, we set $\bar{\kappa}=280$ such that $(\bar{\kappa},\bar\gamma)\in Q$, where $\|E\|_{\max}=2.398\times 10^{-1}$, $\|W\|_{\max}=5.55\times 10^{-1}$, and $\lambda_{\min}(\Psi)=2.58\times 10^{-2}$.
At \textsf{Step~8}, CFRIT was executed with $(\bar{\kappa},\bar{\gamma})$, and the tuned gain was obtained as
\begin{align*}
&F_{\mathcal{E}}^*(\bar\kappa,\bar\gamma) =\\ 
&\ \ \begin{bmatrix}
0.02233808 & 0.27147676 & 0.10718371 & 1.08480513
\end{bmatrix}.
\end{align*}
Meanwhile, the traditional FRIT method yielded
\begin{align*}
&F^*= \\
&\ \ \begin{bmatrix}
0.02238785 & 0.27143312 &0.10715913 & 1.08478539
\end{bmatrix}.
\end{align*}
It is confirmed that the proposed procedure satisfies the specified error tolerance because $\|F_{\mathcal{E}}^*(\bar{\kappa},\bar{\gamma})-F^*\|_2= 3.12 \times 10^{-8}<\epsilon$ holds.

 \begin{figure}
  \begin{subfigure}{.5\textwidth}
  \hspace*{6ex}
  \includegraphics[width=0.75\linewidth]{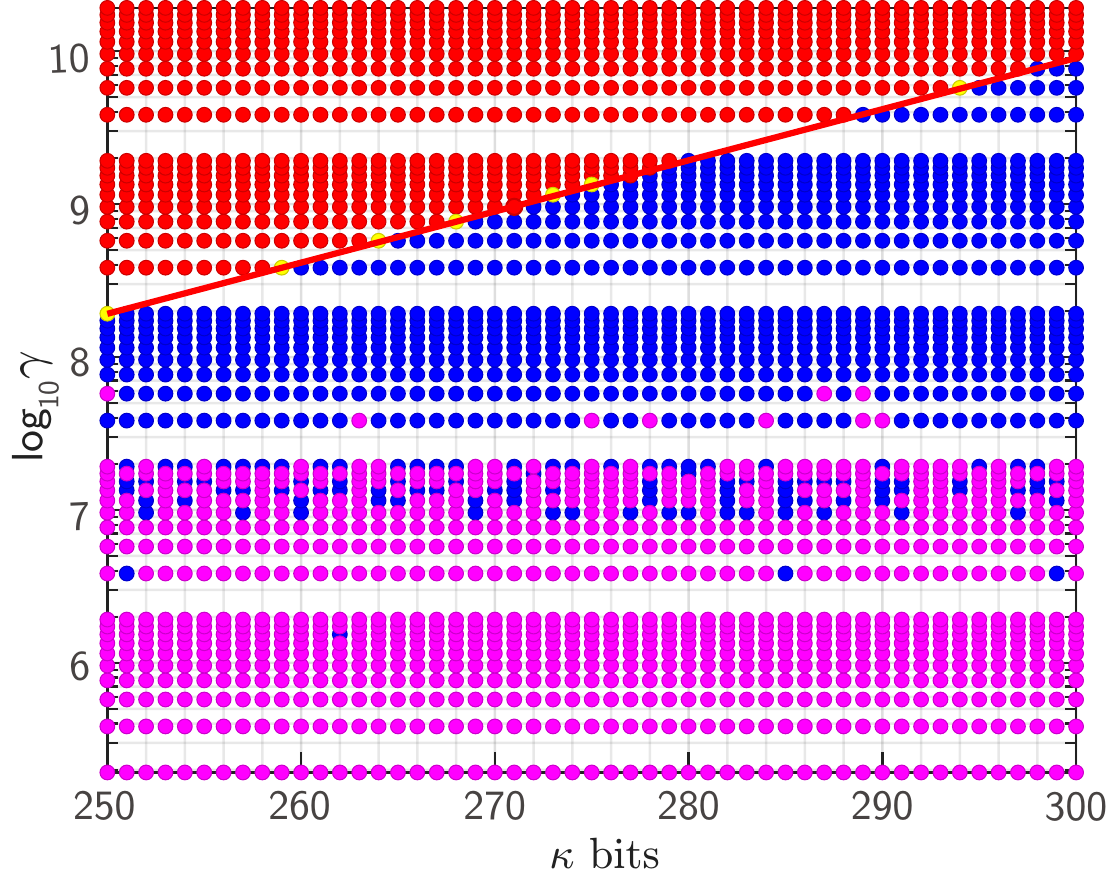}
  \vspace*{1ex}
  \caption{$\kappa-\log_{10}\gamma$ plane}\label{fig:overflow2d}
  \end{subfigure}
  \begin{subfigure}{0.5\textwidth}
  \centering\vspace*{3ex}
  \includegraphics[width=.9\linewidth, keepaspectratio]{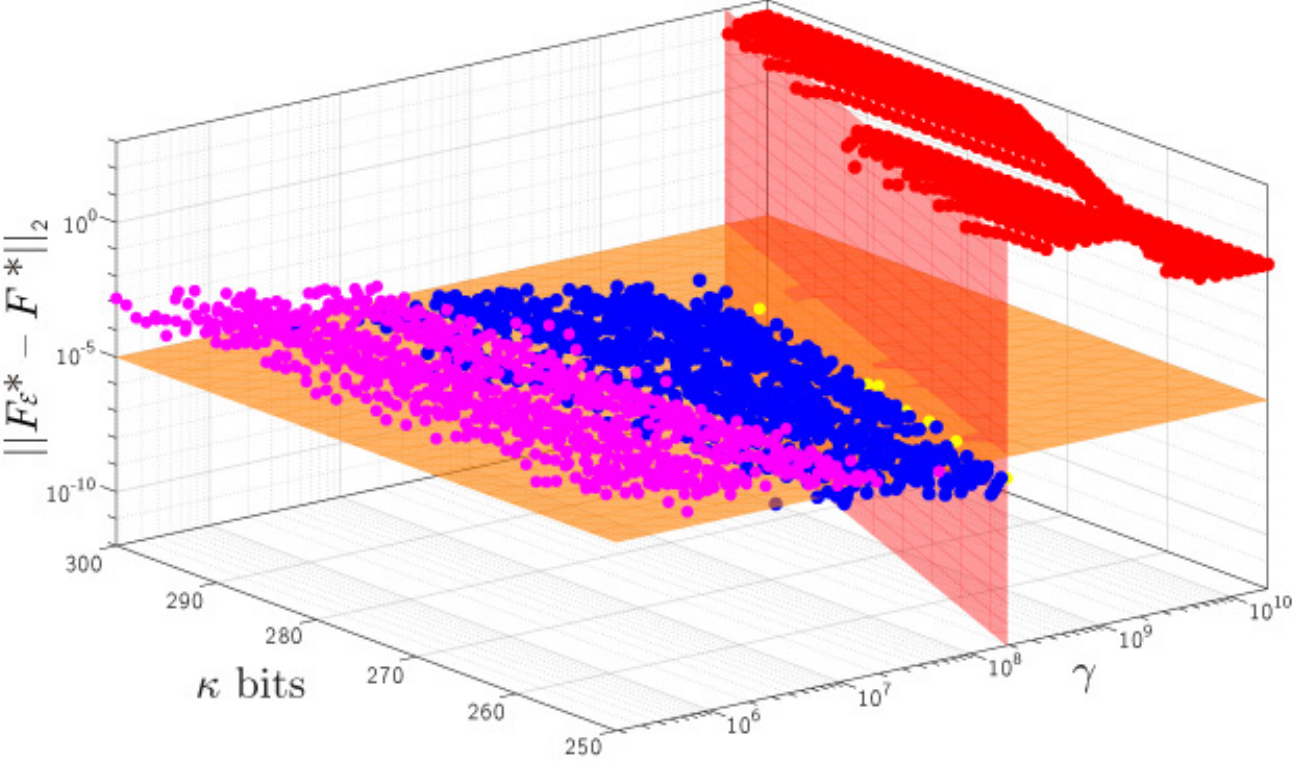}
    \vspace*{2ex}
  \caption{$\|F_{\mathcal{E}}^*(\kappa,\gamma)-F^*\|_2$ in the $(\kappa,\gamma)$ space}\label{fig:overflow3d}
   \end{subfigure}
   \vspace*{-2ex}
 \caption{Quantization-induced error $\|F_{\mathcal{E}}^*(\kappa,\gamma)-F^*\|_2$ and overflow occurrence with respect to $\kappa$ and $\gamma$.
Blue points indicate feasible cases, whereas other colors represent infeasible ones.
Specifically, blue and yellow represent quantization-induced errors below $\epsilon$, while magenta and red correspond to errors below $10^{-4}$ and within the range of $10^0$–$10^2$, respectively.}
\vspace*{2ex}
   \label{fig:overflow}
 \end{figure}

\subsection{Discussion on Admissible Parameter Regions}
The proposed procedure exhibits flexibility in choosing the parameters $(\kappa,\gamma)$, as characterized by the parameter sets $Q$ in~\eqref{ieq:security_parameter_Design} and $\Gamma$ in~\eqref{ieq:quantization_gain_Design}.
This property is examined using the same example.
To investigate it, the security parameter $\kappa$ was varied from 250 to 300 in increments of one, and the quantization gain $\gamma$ was varied from $1.92\times10^{5}$ to $1.92\times 10^{10}$ with equally spaced intervals, where the tolerance $\epsilon$ was fixed to the same value.
For each parameter pair, the CFRIT was executed to verify the presence or absence of overflow and to evaluate the error $\|F_{\mathcal{E}}^*(\kappa,\gamma)-F^*\|_2$.

The results are summarized in Fig.~\ref{fig:overflow}, which illustrates the relationships among $\gamma$, $\kappa$, and the quantization-induced error.
Fig.~\ref{fig:overflow2d} shows the relationship between $\gamma$ and $\kappa$, whereas Fig.~\ref{fig:overflow3d} depicts the parameter space obtained by extending the two-dimensional plane in Fig.~\ref{fig:overflow2d} with an additional axis representing the quantization-induced error.
The blue points denote parameter pairs that satisfy the overflow-free condition and achieve errors below the prescribed tolerance $\epsilon$, while the magenta points also satisfy the overflow-free condition but exhibit errors exceeding $\epsilon$.
The red points correspond to parameter pairs that incur overflows, which consequently yield large errors.
In contrast, the yellow points represent pairs that do not satisfy the overflow-free condition theoretically but remain overflow-free in computation.
Moreover, in Fig.~\ref{fig:overflow3d}, the orange plane represents the prescribed error tolerance $\|F_{\mathcal{E}}^*(\kappa,\gamma)-F^*\|_2=10^{-5}(=\epsilon)$, and the red plane indicates the theoretical boundary of set $Q$, defined by~\eqref{ieq:security_parameter_Design}, which corresponds to the red line in Fig.~\ref{fig:overflow2d}.

In these figures, blue and magenta points are distributed in a mixed manner.
This observation highlights the challenge of defining a clear boundary related to~\eqref{ieq:quantization_gain_Design}.
For future optimization-based parameter design, it is necessary to develop a method that ensures the computational tractability of the parameter determination problem.
Consequently, the visualized results confirm that the proposed conditions effectively identify the feasible parameter region and prevent overflow while maintaining the prescribed accuracy.
These results validate the proposed procedure for parameter choice and indicate its potential extension to optimization-based design.

\section{Conclusion}\label{sec:con}
This study proposed a parameter determination procedure for CFRIT that guarantees overflow-free computation during tuning and ensures that the resulting gain deviates from the traditional FRIT only within a prescribed tolerance. 
The proposed procedure is supported by analytical results on overflow prevention and quantization-induced errors. 
The numerical example confirmed that the proposed method provides reasonable CFRIT parameters as a solution to the parameter determination problem. 
In addition, the admissible parameter region was visualized to demonstrate the flexibility in choosing the parameters.

In future work, we will formulate an objective function to develop an efficient algorithm for finding optimal parameters within the admissible region. 
Furthermore, we plan to conduct a statistical analysis of quantization-induced errors to refine the quantization design and establish a data-driven framework for practical parameter tuning. 
Finally, we will apply a post-quantum cryptographic scheme to CFRIT and further explore the security aspects to establish more secure, encrypted data-driven controller tuning.

\bibliography{ifacconf}             
\end{document}